\documentclass[english]{article}
\usepackage[T1]{fontenc}
\usepackage[latin9]{inputenc}
\usepackage{geometry}
\usepackage{xcolor}
\usepackage{array}
\usepackage{amsmath}
\usepackage{graphicx}
\usepackage{babel}
\usepackage{multirow}
\usepackage{comment}

\title{Many facets of multiparty broadcasting of known quantum information using optimal quantum resource}
\author{Satish Kumar, Anirban Pathak\\ 
Jaypee Institute of Information Technology, A 10, Sector 62, Noida, UP}

\begin{document}

\maketitle
\begin{abstract}
The no-quantum broadcasting theorem which is a weaker version of the nocloning theorem restricts us from broadcasting completely unknown quantum information to multiple users. However, if the sender is aware of the quantum information (state) to be broadcasted then the above restriction disappears and the task reduces to a multiparty remote state preparation. Without
recognizing this fact, several schemes for broadcasting of known quantum states have been proposed in the recent past (e.g., Quantum Inf Process (2017) 16:41) and erroneously/misleadingly referred to as protocols
for quantum broadcasting. Here we elaborate on the relation between the protocols of remote state preparation and those of broadcasting of known quantum information and show that it's possible to broadcast
known quantum information to multiple receivers in deterministic as well as probabilistic manner with optimal resources. Further, the effect of noise on such schemes, and some new facets (like joint broadcasting) of such schemes have been discussed. A proof of principle realization of the proposed optimal scheme using IBM quantum computer is also reported. Possibilities of generalizations of the so-called broadcasting schemes and potential applications are also discussed with appropriate importance.
\end{abstract}

\section{Introduction}
In 1996, a pioneering work of Barnum et al. \cite{BCF+96NoComNoBrod} had established that universal quantum broadcasting is not possible and a set of quantum states is broadcastable if and only if they commute pairwise. In simpler words, noncomuting and thus nonorthogonal states are not broadcastable. This no-go theorem is now known as quantum no-broadcasting theorem \cite{LDX+13FaithfulQB}. This has drawn considerable attention of the community, and has been generalized for every nonclassical finite-dimensional probabilistic theory which satisfies no-signaling criterion \cite{BBL+07GeneralizedNoBrodThm}. It may be argued that since unknown quantum states cannot be cloned so they cannot be broadcasted and consequently no-broadcasting is a trivial consequence of the nocloning theorem. However, a closer look into the situation would reveal that no-broadcasting is a weaker version of the nocloning theorem. To be specific, the nocloning theorem implies that there does not exist any CPTP map $\cal E$ such that for any pair of nonorthogonal pure states $\rho_{i}, i\in\{1,2\}$ it can produce $\cal E$ $(\rho_{i})=\rho_{i}\otimes \rho_{i}$ $\forall i$. Whereas separability of the final state is not required in case of the broadcasting, which requires a CPTP map $\cal E^{\prime}$ which can broadcast a state $\rho$ on $\cal H$ to $\cal H_{A}\otimes H_{B}$ iff $Tr_{A}$($\cal E^{\prime}(\rho))$=$Tr_{B}$(${\cal E^{\prime}}(\rho))=\rho$. Absence of such a CPTP map $\cal E^{\prime}$ for unknown quantum states (or nonorthogonal quantum states) is referred to as quantum no-broadcasting theorem \cite{BBL+07GeneralizedNoBrodThm}.

Despite the strong presence of the quantum no-broadcasting theorem, a set of works have recently appeared which claim to have performed quantum broadcasting \cite{LDX+13FaithfulQB,GH07QsharedB,YZL17,ZT18,yu2018general}. Specifically, in Ref. \cite{YZL17}, Yu et al. proposed a scheme that enabled sender Alice to broadcast an arbitrary single-qubit state to two receivers (Bob and Charlie) using a four-qubit cluster state. Interestingly, in their protocol, Alice had complete knowledge of the state to be broadcast. So it was essentially, a task to remotely prepare multiple copies of a known quantum state. Neither noclononing theorem nor quantum no-broadcasting theorem prohibits such a task, but it was slightly misleading or erroneous to refer to such a protocol as a protocol for quantum broadcasting (without a clear mention of the fact that the state to be broadcasted is known), especially in view of the quantum no-broadcasting theorem. Similar misleading nomenclature was also used in \cite{LDX+13FaithfulQB,BBL+07GeneralizedNoBrodThm,ZLY20,WLZ08RealizeQBnMultiSign,XL16QBnBlind,ZQZ16Improvement,YZ18GeneralQB}. Later on, Yu et al. scheme was modified in Ref. \cite{ZT18} to obtain a probabilistic version of the original scheme by using a non-maximally entangled cluster state as a quantum channel. Keeping this in mind, in rest of the paper, by boradcasting of quantum information we would mean broadcasting of known quantum information unless stated otherwise.  In view of the above, here we aim to clearly state that only a known quantum state can be broadcasted and the same can be done in an optimal way. Specifically, an optimal scheme for the same is proposed here and it's established that the task is equivalent to multiparty joint remote state preparation. Further, the scheme is realized using IBM quantum computer. Before we proceed further, it would be apt to note that the optimality mentioned above and to be discussed in the rest of the paper is for the broadcasting of a general class of quantum states (say a qubit in general) implemented with the help of entangled states comprised of qubits only (i.e., higher dimensional states are not used). This point needs specific attention as the amount of ebit used in the recently proposed broadcasting protocol of Hillery et al. (cf.   Ref.\cite{hillery2022broadcast}) is less than that in our scheme. However, their scheme is designed for broadcasting of a restricted class of quantum states (e.g., a qubit of the form $|\psi(\theta)\rangle =\frac{\exp{(\iota \theta})|0\rangle+\exp{(-\iota \theta)}|1\rangle}{\sqrt{2}}$  and the quantum channel used for the three party broadcasting involved a shared entangled state made of two qubits which are with receivers and a qutrit, which is with the sender. Here, it would be apt to note that the work of Hillery et al., was clear on the fact that what we need to do to implement a broadcasting protocol is essentially remote state preparation as stressed above. The benefit of a common remote state preparation is that it requires a lesser amount classical communication in comparison to teleportation. Interestingly, the capability of teleportation ensures the capability of remote state preparation as if we can teleport an unknown state we must be able to teleport a known one. Now, a restricted class of quantum states can be teleported with a lesser amount of classical resources. This fact was exploited in the work of Hillery et al. Such an approach along with the use of higher dimensional states can be used to further reduce the resource requirement, but such an approach will neither be general nor easily realizable with the available technologies. Keeping this in mind we restricted ourselves to the broadcasting of a qubit in general using entangled states made of qubits only.

The rest of the paper is organized as follows. In Section \ref{sec:Broadcasting-of-known}, the proposed schemes of quantum broadcasting have been discussed using optimal quantum resources. The possible generalization using optimal resources have been made for different facets of quantum broadcasting in Section \ref{sec:Possible-generalizations-and}. The effect of different types of noises has been studied on a pair of Bell state and a cluster state in Section \ref{sec:Effect-of-noise}. Further, the proposed scheme has been experimentally realized in IBM Quantum in Section \ref{sec:Realization-of-the}. Finally, the paper is concluded in Section \ref{conclusion}. 

\section{Broadcasting of known quantum information to multiple users by using
optimal resources\label{sec:Broadcasting-of-known}}

Let us discuss a two party quantum broadcast scheme introduced by Yan et al. \cite{YZL17}, where Alice want to broadcast a message $\alpha|0\rangle+\beta|1\rangle$ such that $|\alpha|^{2}+|\beta|^{2}=1$ to
Bob and Charlie. For that Alice prepares a four qubit cluster state given by 

\begin{equation}
|\phi\rangle_{1234}  = \frac{1}{2}(|0000\rangle+|0101\rangle+|1010\rangle-|1111\rangle.
\end{equation}

Alice keeps particle $1$ and $2$ with herself and sends particle $3$ and $4$ to Bob and Charlie, respectively. Now, Alice choose a measurement basis set $\{|\varphi_{1}\rangle,|\varphi_{2}\rangle,|\varphi_{3}\rangle,|\varphi_{4}\rangle\}$
where

\begin{equation}
\begin{split}
|\varphi_{1}\rangle =  \alpha^{2}|00\rangle+\alpha\beta|01\rangle+\alpha\beta|10\rangle-\beta^{2}|11\rangle\\
|\varphi_{2}\rangle = \alpha\beta|00\rangle-\alpha^{2}|01\rangle+\beta^{2}|10\rangle+\alpha\beta|11\rangle\\
|\varphi_{3}\rangle = \alpha\beta|00\rangle+\beta^{2}|01\rangle-\alpha^{2}|10\rangle+\alpha\beta|11\rangle\\
|\varphi_{4}\rangle = \beta^{2}|00\rangle-\alpha\beta|01\rangle-\alpha\beta|10\rangle-\alpha^{2}|11\rangle
\end{split}
\end{equation}

and measures particles $1$ and $2$ in this measurement basis. Finally,
Bob and Charlie can retreive the message by applying appropriate unitaries
depending on the Alice's measurement outcomes on their corresponding
particles. This way Alice broadcasts a message to Bob and Charlie.
Similar excercise has been done in a paper of Yun-Jing Zhou et al.
\cite{ZT18} where they showed probabilistic quantum broadcasting which uses a four qubit non-maximally entangled state. When different messages is sent to different receivers then this scheme
is called multi-output quantum teleportation. These schemes have been
generalized in some recent papers \cite{LDX+13FaithfulQB,YZ18GeneralQB}. However it is to be noted that the message sent by Alice is known to her and hence the scheme can be viewed as a scheme for multi-party remote state preparation. Also the
resources used to broadcast two party is a four qubit cluster state.
As discussed in reference \cite{sisodia2017design}, the teleportation
of a quantum state having $m-$unknown coefficient requires $\lceil\log_{2}m\rceil$
Bell states. In the same line, broadcasting an information in form
of quantum state having $m-$coefficient to $n-$receivers would require
$\lceil\log_{2}m^{n}\rceil$ Bell states. So, the example discussed
above requires $\lceil\log_{2}4=2\rceil$ Bell state instead of four
qubit entangled state.

Let $|\psi^{+}\rangle_{13}$ and $|\psi^{+}\rangle_{24}$
are the two Bell state where $|\psi^{+}\rangle=\frac{1}{\sqrt{2}}(|00\rangle+|11\rangle)$.
Alice keeps first particle of each Bell state with herself and sends
other particle to Bob and Charlie. Alice sends a message to Bob and
Charlie at their ends using the process called remote state preparation
(RSP) \cite{bennett2001RSP,pathak2013book}. Readers might get confused
about the simultaneous transfer of message to each receivers. But
it is to be noted that to get the desired message in RSP process,
sender has to send one classical bit of information according to which
receiver will apply unitary. These $1-$bit of classical messages
is sent to each receiver simultaneously over the conference call
so that the receivers can apply unitaries and get the desired message
simultaneously. The optimal state corresponding to resources used
for different purposes has been reported in Table \ref{tab:TableOptimize}.
A general method for selecting quantum channels for various schemes
of controlled quantum communication has been reported by some of the present authors in Ref \cite{thapliyal2015general}.
In a similar manner, the possible generalizations for different
variants of quantum broadcasting have been done in the next section. 

\begin{table}
\centering
\begin{tabular}{|>{\centering}m{1cm}|>{\centering}m{8cm}|>{\centering}m{4cm}|c|c|}
\hline 
S.No & \centering{State used as channel for broadcasting} & Purpose & Party & Optimal State \\
\hline 
1 & $\frac{1}{2}(|0000\rangle+|0101\rangle+|1010\rangle-|1111\rangle)_{A_{1}A_{2}B_{1}B_{2}}$\cite{YZL17} & Quantum broadcasting & 3 & $|\phi^{+}\rangle_{A_{1}B_{1}}\otimes|\phi^{+}\rangle_{A_{2}B_{2}}$\\
\hline
2 & $\frac{1}{2}(|0001\rangle+|0110\rangle+|1011\rangle+|1100\rangle)_{A_{1}A_{2}B_{1}B_{2}}$\cite{LDX+13FaithfulQB} & Quantum broadcasting & 3 & $|\phi^{+}\rangle_{A_{1}B_{1}}\otimes|\phi^{+}\rangle_{A_{2}B_{2}}$\\
\hline
3 & $\frac{1}{4\sqrt{2}}(|00000000000000000000\rangle+|00000000010000000001\rangle+...+|11111111101111111110\rangle+|11111111111111111111\rangle)$\cite{ZK21Efficient} & Circular quantum broadcasting & 5 & $|\phi^{+}\rangle_{A_{i}B_{i}}^{\otimes10}$\\
\hline 
\end{tabular}
\caption{The optimal quantum states corresponding to the previously reported quantum states as channels for quantum broadcasting. Here, $|\phi^{+}\rangle=\frac{1}{\sqrt{2}}|00\rangle+|11\rangle$.}
\label{tab:TableOptimize}
\end{table}

\section{Possible generalizations and potential applications \label{sec:Possible-generalizations-and}}
A possible generalization for teleporting $n-$qubit state with $m-$unknown coefficient has been made in the recent past \cite{sisodia2017design}. Also, a general method has been developed for selecting a quantum channel for bidirectional controlled state teleportation (BCST) and other schemes for controlled quantum communication \cite{thapliyal2015general}. In the previous section, we have discussed that quantum broadcasting can be seen as multiparty RSP. A pre-shared entangled state is required to perform RSP. As generation and maintenance of multipartite entangled states are difficult, we may restrict ourselves to a situation where the multiparty RSP task is to be implemented with the help of bi-partite entangled states only. In this situation, we would require $m$ pre-shared entangled state for $m$ remote parties (receivers). Thus, a generalized channel for quantum broadcasting can be given as

\begin{eqnarray}
|\chi\rangle & = & \otimes_{i=1}^{m}|\phi_{2i-1,2i}^{i}\rangle,\label{eq:General broadcasting},
\label{eq:General broadcast}
\end{eqnarray}
where $|\phi^{i}\rangle$ such that $|\phi^{i}\rangle$ is not essentially equal to $|\phi^{j}\rangle$ is a two-qubit entangled state for broadcasting, and qubits $\{2i-1\}$ are with Alice and qubits $\{2i\}$ are with $n$ receivers. If $|\phi^{i}\rangle$ is maximally entangled Bell state then Eqn. (\ref{eq:General broadcast}) is optimal generalized channel for deterministic perfect quantum broadcasting and if $|\phi^{i}\rangle$ is non-maximally entangled Bell state then
Eqn. (\ref{eq:General broadcast}) is a generalized optimal channel for probabilistic or imperfect quantum broadcasting.
One can check the quantum channel used in Ref. \cite{ZT18} for probabilistic quantum broadcasting is a special case of our proposed general channel. Previously reported quantum channels for various schemes of broadcasting can be replaced by the generalized optimal channel proposed above (see Table \ref{tab:TableOptimize}). However, if we consider higher dimensional states (qudits), useful features of broadcasting can be realized with a lesser amount of e-bits. For example, in Ref. \cite{hillerry2023broadcasting}, a scheme for broadcasting of a restricted class of states to multiple receivers has been proposed. As noted in the previous section, we are restricting ourselves to qubit based implementations, as they are easy to realize and as proof of principle experiments can be performed with IBM quantum computers. 

In our view, quantum broadcasting is a variant of remote state preparation. As there are many variants of remote state preparation (see \cite{SSB+15} references therein for details) there can be many variants of quantum broadcasting of known quantum information. For example, in analogy with deterministic and probabilistic joint remote state preparation,  probabilistic and deterministic controlled (and without controller) bidirectional remote state preparation, we can construct schemes for deterministic and probabilistic joint quantum broadcasting,  probabilistic and deterministic controlled (and without controller) bidirectional quantum broadcasting. Though such variants of broadcasting are not yet explored with specific attention, they are obvious extensions of the general schemes for remote state preparation discussed by some of the present authors in Ref. \cite{SSB+15}. Here, it would be apt to note that in joint remote state preparation, usually there are two senders. One of them (say Alice 1) knows the amplitude information, whereas the other sender (say Alice 2) knows the phase information. Specifically, if we think that a qubit $\cos(\theta)|0\rangle+\exp(\iota \phi)|1\rangle$ is to be broadcasted, but Alice 1 only knows $\theta$ and Alice 2 knows $\phi$, then they can together prepare the state remotely. An analogous scheme for joint broadcasting will have 2 senders, each with incomplete information about the state to be broadcasted. The simplest quantum resource that can be used to implement such a scheme is a product of $m$ pieces of 3 qubit entangled states, if there are $m$ receivers and it's assumed that a single qubit is to be broadcasted.  For the 2 senders and $m$ receivers case, a generalized channel for joint quantum broadcasting can be given as
\begin{center}
\begin{eqnarray}
|\varphi\rangle & = & \otimes_{i=1}^{m}|\phi_{3i-2,3i-1,3i}^{i}\rangle.\label{eq:Joint broadcasting}
\label{eq: joint broadcasting}
\end{eqnarray}
\par\end{center}
If $|\phi^{i}\rangle$ is maximally (non-maximmaly) entangled three qubit state then Eqn. (\ref{eq: joint broadcasting}) is the generalized channel for joint (probabilistic joint) quantum broadcasting. This is consistent with the existing schemes for remote state preparation, and easily implementable, but it's not optimal and if we try to generalize it to $n$ sender case, then the analogous requirement will be of $(n+1)$ qubit entangled state, which is difficult to produce and maintain for relatively large values of $n$. Keeping this in mind, let's look at the possibilities of quantum broadcasting of known information by $n$ senders to $m$ receivers.  In such a situation, the information about the known quantum state is to be shared among $n$ senders, and we may assume that the senders are referred to as Alice 1, Alice 2,...,Alice $n$. Now, to broadcast, $\cos(\theta)|0\rangle+\exp(\iota \phi)\sin(\theta)|1\rangle$, information about $\theta$ and $\phi$ are to be distributed among the senders, and for convenience, we may assume that Alice 1 knows $\theta$ and Alice $j:j\in{2,3,...,n}$ knows $\phi_{j-1}:\phi_{1}+\phi_{2}+\phi_{3}+...\phi_{n}$. In such, a situation, senders will sequentially encode the information available to them through appropriate rotation. There are various options for the implementations. One is to prepare $m$ copies of $(n+1)$ partite entangled state, another simple approach would be as follows. Only Alice $n$ shares entangled state(s) with the receivers. Alice 1 prepares $\cos(\theta)|0\rangle+\sin(\theta)|1\rangle$  and sends that to Alice 2 who applies $P(\phi_{1})$ gate to it and transforms the state to $\cos(\theta)|0\rangle+\exp(\iota \phi_{1})\sin(\theta)|1\rangle$ and sends that to Alice 3 who applies $P(\phi_{2})$ to encode the state information available to her and sends to next sender. The process continues till Alice $n$ encodes the information available to her and obtains $\cos(\theta)|0\rangle+\exp(\iota \phi)\sin(\theta)|1\rangle$ which she can subsequently broadcast to $m$ receivers using $m$ number of two-qubit shared entangled states as described above. Such a general scheme for $n$ senders and $m$ receivers have been discussed in Ref. \cite{hillerry2023broadcasting}, but the quantum resources required there involve qudits and are much more complicated. 

The controller in controlled quantum broadcasting schemes can control which receivers will get the message sent by the sender (Alice). Until the controller reveals his/her measurement outcome, receivers will not be able to extract the message. The generalized channel for controlled quantum broadcasting can be given in analogy to the channel for controlled quantum remote state preparation as

\begin{eqnarray}
|\tau\rangle & = & \sum_{k=1}^{m}|\chi\rangle_{k}|a_{k}\rangle\label{eq:controlled broadcasting},
\end{eqnarray}
where {$|a_{k}\rangle$} is a set of $m$ single qubit  mutually orthogonal states and $|\chi\rangle_{k}\neq|\chi\rangle_{k'}$, where $|\chi\rangle$ is defined in Eqn. (\ref{eq:General broadcast}). Interestingly, this can be easily generalized to $n$ senders and $m$ receivers case following the strategy described above, and in addition to that for implementing a controlled broadcasting scheme with only two qubit entnagled states, the controller would require to prepare $m$ two qubit entangled states randomly (say $m$ Bell states are prepared randomly) and share that in a manner that first qubit of the $i^{th}$ entangled state is with Alice $n$ and the second qubit is with Bob $i$, but the controller keeps the information about which entangled state she has shared among Alice $n$ and Bob $i$ secretly. Until this information is disclosed, broadcasting will not be successful. Thus, in short, it's possible to design several variants of quantum broadcasting of known information, and all of them can in principle be implemented using two-qubit entangled states. In general, with some tricks, the state described in Eqn. (\ref{eq:General broadcast}) can be used as the optimal channel for every variant of broadcasting task. We neither require complicated multi-qubit states used in \cite{LDX+13FaithfulQB,YZL17,ZT18,ZK21Efficient}  nor we require higher dimensional states used in  \cite{hillery2022broadcast, hillerry2023broadcasting}. 

In the above, we have mentioned of bidirectional remote state preparation and its analogous schemes called bidirectional boradcasting, which allows both Alice and Bob to broadcast information. Now, in two party scenario involving only one Alice and one Bob, it's essentially bidirectional as there are two directions only. However, if we think of $n$ senders and $m$ receivers in general and allow both way broadcasting (without considering a joint preparation of the state to be broadcasted), we will have $2(m\times n)$ direction of communication, and we will have multi-directional quantum broadcasting. In a more general scenario, we may consider that there are $n$ parties and each party can broadcast an known quantum state to all the other parties present in the quantum network. The notion of senders and receivers becomes vague in this situation and we have $n(n-1)$ direction of communication. Each such communication channel would require a Bell type two qubit entangled state and the  generalized channel for multi-directional quantum broadcasting can be given as
\begin{eqnarray}
 |\kappa\rangle & = & \otimes_{i=1}^{(n\times(n-1))}|\phi_{2i-1,2i}^{i}\rangle  
\label{eq:multi-direction broadcasting}.   
\end{eqnarray}
Obviously, this state is experimentally realizable and it's a simple exercise to construct schemes for very general situations like deterministic and probabilistic controlled multidirectional broadcasting and then to construct special cases of that.

So far, we have primarily discussed different possible variants of quantum broadcasting and the tricks that can be used to implement them with optimal resources or simply using the required number of copies of Bell states or Bell type states. But a question remained, why should we do that, or simply, why will someone be interested in performing broadcasting? In our day-to-day life, we know several applications of classical broadcasting. In addition, in a set of recent works \cite{hillery2022broadcast, hillerry2023broadcasting}, it's explicitly shown that quantum broadcasting can be used for realizing schemes for quantum voting, quantum identity authentication and quantum cryptography. Further, it's shown to be relevant in the context of measurement based quantum computing.

\section{Effect of noise \label{sec:Effect-of-noise}}

Quantum states are being distributed among different parties whenever any quantum communication protocol is realizes. Like quantum channel used
in quantum broadcasting protocols are distributed among sender and
receivers. The distribution of quantum states are affected by different
noises present in the environment. So, it is necessary to study the
effect of noise on the quantum channel used for communication. Any
quantum communication protocol is considered to be useful only if
it is robust in the presence of noise (for details see \cite{banerjee2017Noise,banerjee2018Noise,sharma2016Noise,satish2022experimentalNoise,thapliyal2018Noise}).
Evolution of any quantum state in the presence of noise can be described
by a CPTP (complete positive trace preserving) map, which maps an initial
state $\rho$ to a final state $\rho'$. This type of evolution is
mathematically represented as

\begin{eqnarray}
\rho' & = & \sum_{i}K_{i}\rho K_{i}^{\dagger},
\end{eqnarray}

where $K_{i}^{'}$s are the Kraus operator which follows the completness
relation $\sum_{i}K_{i}K_{i}^{\dagger}=1$. Some of the most relevent
noises which are usually considered in any quantum communication schemes
are bit-flip, phase-flip, amplititude damping and depolarizing noise.
The details of these noises are explained as follows.

Bit flip noise flips qubit $|0\rangle$ to $|1\rangle$ with certain probability $p$ and vice-versa. The Kraus operator corresponding to the bit
flip noise channel is expressed as
\begin{equation}
    K_{0}^{BP}=\sqrt{p}
    \begin{pmatrix} 
    1 & 0 \\
    0 & 1
    \end{pmatrix},
    \qquad
    K_{1}^{BP}=\sqrt{1-p} 
    \begin{pmatrix} 
    0 & 1 \\
    1 & 0
    \end{pmatrix}.
\end{equation}

Depolarizing noise in quantun channel converts a pure quantum state into a maximally mixed state. The Kraus operator corresponding to the depolarizing noise channel is expressed as
\begin{equation}
   K_{0}^{DP}=\sqrt{1-\frac{3p}{4}}
    \begin{pmatrix} 
    1 & 0 \\
    0 & 1
    \end{pmatrix},
    \qquad
    K_{1}^{DP}=\frac{\sqrt{p}}{2}
    \begin{pmatrix} 
    0 & 1 \\
    1 & 0
    \end{pmatrix},
    \qquad
    K_{2}^{DP}=\frac{\sqrt{p}}{2}
    \begin{pmatrix} 
    0 & -i \\
    i & 0
    \end{pmatrix},
    \qquad
    K_{3}^{DP}=\frac{\sqrt{p}}{2}
    \begin{pmatrix} 
    1 & 0 \\
    0 & -1
    \end{pmatrix},
\end{equation}
where $p$ is the probability with which a pure quantum state is depolarized.

Amplitude damping noise in quantum channels causes spontaneous emission of photons. The Kraus operator corresponding to the amplitude damping noise channel is expressed as
\begin{equation}
    K_{0}^{AD}=
    \begin{pmatrix} 
    1 & 0 \\
    0 & \sqrt{1-p}
    \end{pmatrix},
    \qquad
    K_{1}^{AD}= 
    \begin{pmatrix} 
    0 & \sqrt{p} \\
    0 & 0
    \end{pmatrix},
\end{equation}
where $p$ represents the amplitude damping probability which ranges from 0 to 1.

Phase damping noise in quantum channel losses relative phase information of quantum state in the channel, but preserves energy of the state. The Kraus operator corresponding to the phase damping noise channel is expressed as
\begin{equation}
    K_{0}^{PD}=\sqrt{1-p}
    \begin{pmatrix} 
    1 & 0 \\
    0 & 1
    \end{pmatrix},
    \qquad
    K_{1}^{PD}=\sqrt{p} 
    \begin{pmatrix} 
    1 & 0 \\
    0 & 0
    \end{pmatrix},
    \qquad
    K_{2}^{PD}=\sqrt{p} 
    \begin{pmatrix} 
    0 & 0 \\
    0 & 1
    \end{pmatrix},
\end{equation}
where $p$ represents the phase damping probability.

\begin{figure}
\centering
\includegraphics[scale=0.8]{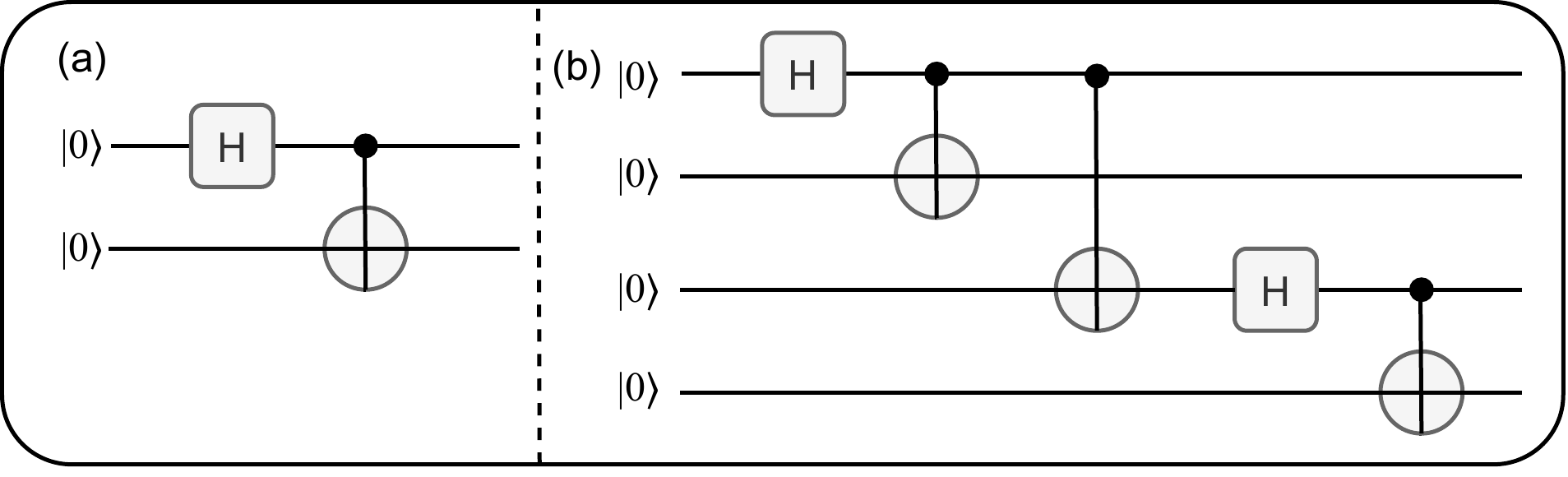}
\caption{Quantum circuit for preparation of (a) Bell state and (b) cluster state.}
\label{Circuit preparation}
\end{figure}

\begin{figure}
\centering
\includegraphics[scale=0.8]{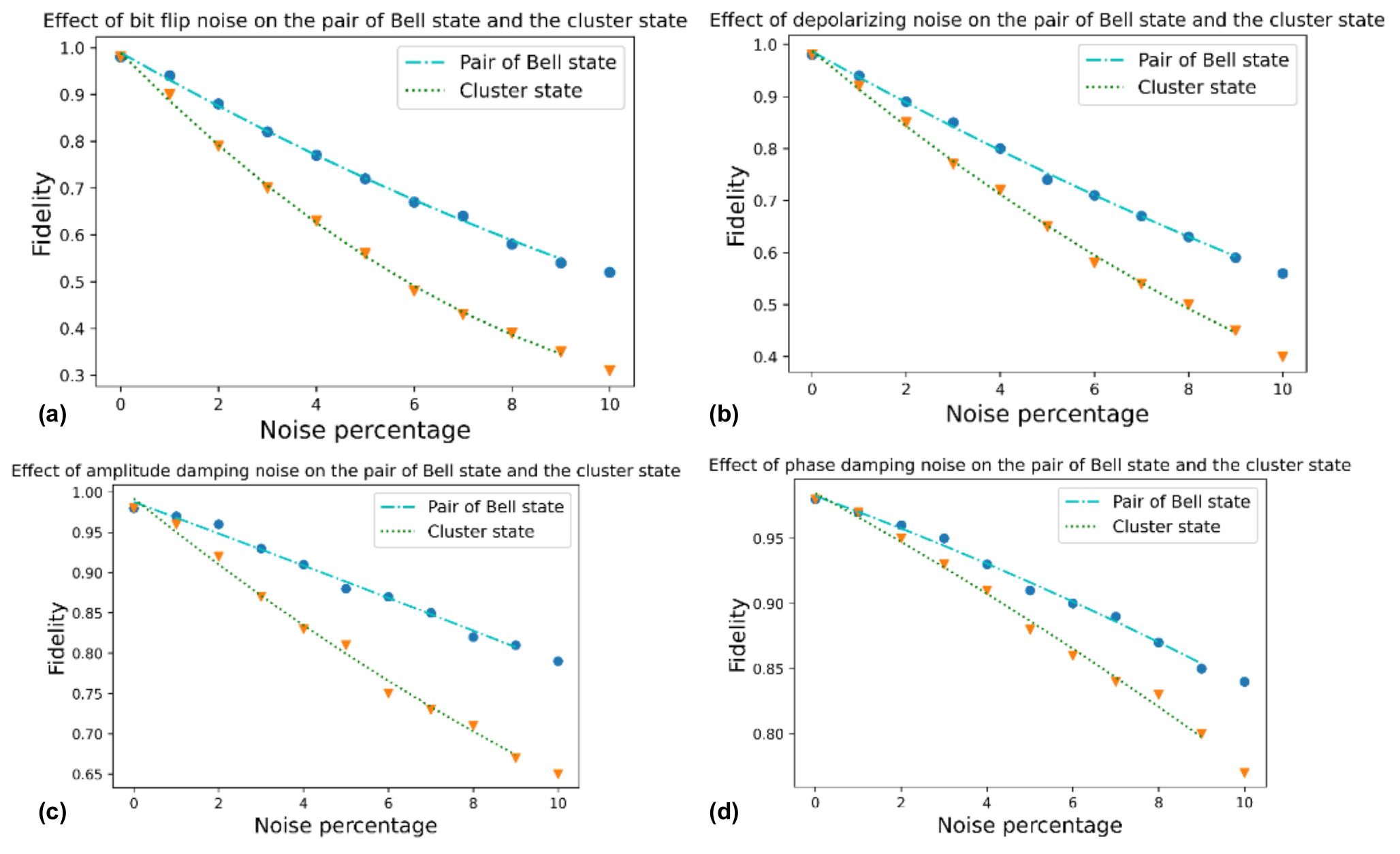}
\caption{Effect of (a) bit-flip (b) depolarizing (c) amplitude damping (d) phase damping noise on the pair of the Bell state and the cluster state.}
\label{Plots}
\end{figure}

Here, the effect of noise has been studied using noise model building technique. This technique is available on qiskit \cite{qiskit.noise}, an open source software to work with quantum computer. The quantum circuits used for preparation of Bell state and cluster state are shown in Fig. \ref{Circuit preparation}. $H$ represents Hadamard operation, which transforms states in computational basis to states in diagonal basis and vice versa. Different type of noises are introduced on these circuits and fidelity is studied after measuring output of these circuits before and after introducing noise. Note that the output of two EPR circuit (a quantum circuit shown in Fig. \ref{Circuit preparation}(a)) has been compared with the output of cluster state circuit (a quantum circuit shown in Fig. \ref{Circuit preparation}(b)). Further, in view of the fact that task done in Ref. \cite{YZL17} using cluster state can be done using two Bell states, it is observed that pair of Bell state is performing better in presence of each noise listed here compare to cluster state. The performance of quantum channel or states is shown in plots of fidelity measure versus noise percentage which is shown in Fig. \ref{Plots}. 

\section{Realization of the proposed scheme using IBM quantum computer \label{sec:Realization-of-the}}
We have seen that quantum broadcasting to two receivers can be done using a cluster state or two Bell states. Broadcasting with two Bell states is an optimal resource and is more robust against noise than with a cluster state (cf. Fig. \ref{Plots}). In this section, a quantum circuit is designed where a quantum state $\alpha|0\rangle + \beta|1\rangle,$ 
$\forall$ $|\alpha|^{2}=|\beta|^{2}=\frac{1}{2}$ is broadcasted to two receivers. 

\begin{figure}
\centering
\includegraphics[scale=0.8]{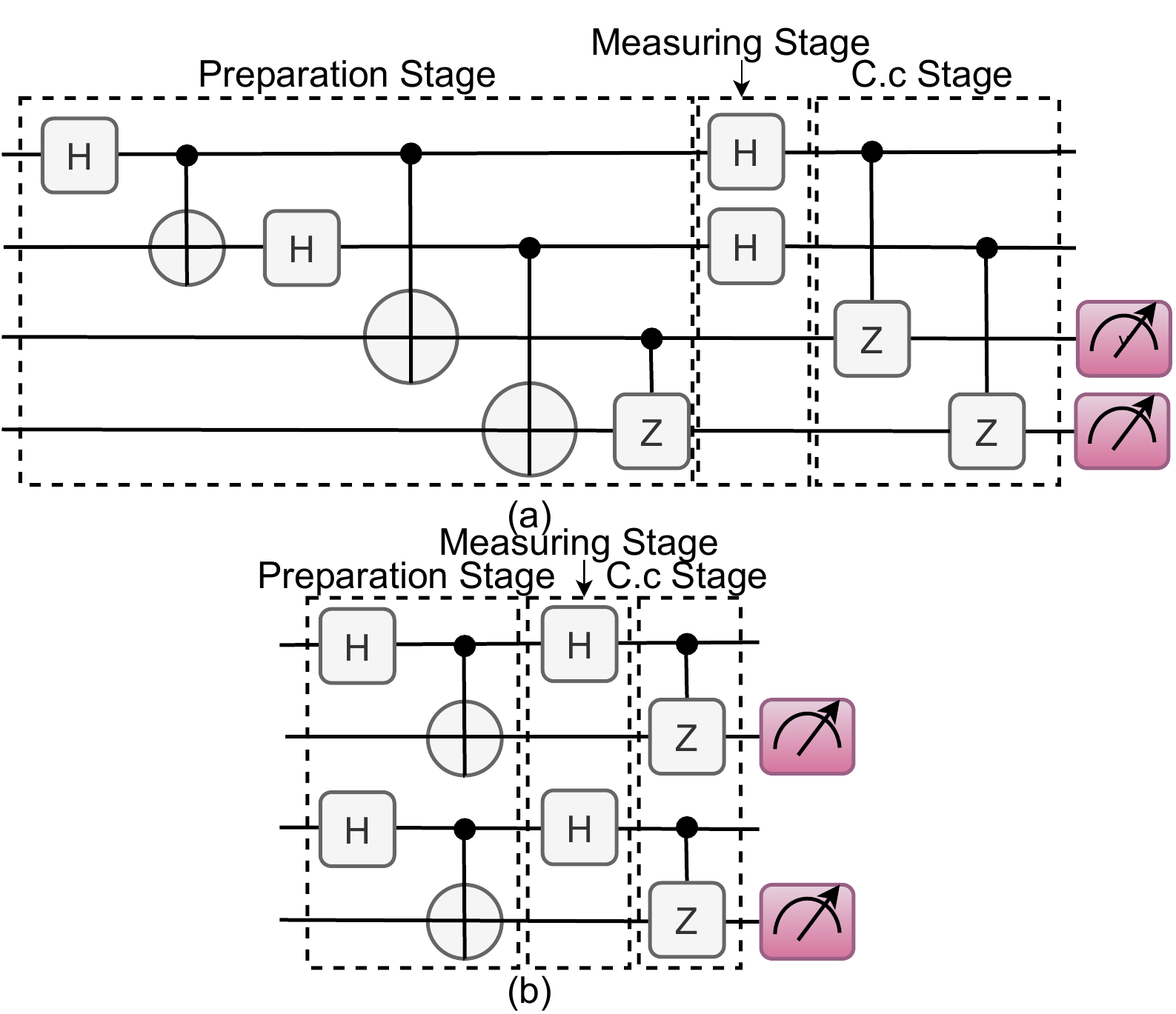}
\caption{Quantum circuit to broadcast a state $\alpha|0\rangle + \beta|1\rangle$ with $|\alpha|^{2}=|\beta|^{2}=\frac{1}{2}$ to two receivers using (a) a cluster state and (b) two Bell state. Here, C.c stands for classical communication.}
\label{Comparative qcircuit}
\end{figure}

While doing broadcasting with a cluster state, Alice first prepares a cluster state $\frac{1}{2}(|0000\rangle +|0101\rangle +|1010\rangle +|1111\rangle)$ (cf. leftmost block of Fig. \ref{Comparative qcircuit}(a)) and then measure her first and second qubit in the basis $\{|++\rangle + |+-\rangle +|-+\rangle +|--\rangle\}$ where $|\pm \rangle=\frac{1}{\sqrt{2}}(|0\rangle \pm |1\rangle)$ (cf. middle block of Fig. \ref{Comparative qcircuit} (a)) and send the last two qubit to respective receivers. Finally, the two receivers apply an appropriate unitary to their corresponding qubits depending on the Alice's measurement outcomes. Here, the two receivers apply the pauli Z operation  on their corresponding qubits if Alice's measurement outcome is 11 (cf. rightmost block of Fig. \ref{Comparative qcircuit} (a)). In this way, the two receiver will obtain the state $\alpha|0\rangle + \beta|1\rangle,$ $\forall$ $|\alpha|^{2}=|\beta|^{2}=\frac{1}{2}$.  

While doing broadcasting with the Bell state, Alice first prepare the two Bell states $\frac{1}{\sqrt{2}}(|00\rangle +|11\rangle)^{\otimes{2}}$ (cf. leftmost block of Fig. \ref{Comparative qcircuit}(b)) and then measure her first qubit of each pair in the $\{|+\rangle, |-\rangle\}$ basis (cf. middle block of Fig. \ref{Comparative qcircuit} (b)) and send other qubit to respective receivers. Finally each receivers apply an appropriate unitary on their corresponding qubits depending on the Alice's measurement outcome. Here, each receiver apply pauli Z operation if the outcome of Alice's qubit of the corresponding shared Bell state is 1 (see rightmost block of Fig. \ref{Comparative qcircuit} (b)). In this way, the two receivers will obtain the desired state $\alpha|0\rangle + \beta|1\rangle,$ $\forall$ $|\alpha|^{2}=|\beta|^{2}=\frac{1}{2}$.

The circuit shown in Fig. \ref{Comparative qcircuit} is experimentally executed on ibmq\_manila which is a 5-qubit quantum computer based on Falcon processor and whose quantum volume (QV) \cite{CBS+19} is 32 which is the maximum among the five qubit quantum computers available on IBM Quantum. The QV can be viewed as a benchmark for a quantum computer indicating the size of the quantum circuits that can be reliably run on that quantum computer \cite{CBS+19, JJB+21}. More precisely, QV quantifies the greatest random circuit of equal width and depth that a quantum computer can successfully implement. Naturally, QV is sensitive to various kinds of possible errors and device imperfections (system error rates) and the greater is the QV the better is the computer. One can access the available quantum computers on IBM Quantum based on one's requirement \cite{ibmq}. Here, ibmq\_manila is chosen for executing the circuit based on its topology, quantum volume and availability. The calibration data of ibmq\_manila is shown in Table \ref{tab:Calibration-data-of} which shows decoherence time of qubits and other errors related to it. 

\begin{table}
\begin{centering}
\begin{tabular}{|c|c|c|>{\centering}p{2cm}|>{\centering}p{2cm}|>{\centering}p{2cm}|>{\centering}p{5cm}|}
\hline
Qubit & T1 ($\mu s$) & T2 ($\mu s$) & \centering{}Frequency (GHz) & \centering{}Readout assignment error & \centering{}Single-qubit Pauli-X-error & \centering{}CNOT error\tabularnewline
\hline 
$Q_{0}$ & 110.78 & 115.99 & \centering{}4.96 & \centering{}$2.61\times10^{-4}$ & \centering{}$2.73\times10^{-4}$ & \centering{}cx0\_1: $7.84\times10^{-3}$\tabularnewline
\hline 
$Q_{1}$ & 197.24 & 70.83 & \centering{}4.84 & \centering{}$4.63\times10^{-2}$ & \centering{}$2.12\times10^{-4}$ & cx1\_0: $7.84\times10^{-3}$, cx1\_2: $1.30\times10^{-2}$
\tabularnewline
\hline 
$Q_{2}$ & 153.43 & 25.39 & \centering{}5.04 & \centering{}$1.99\times10^{-2}$ & \centering{}$4.85\times10^{-4}$ & \centering{}cx2\_1: $1.30\times10^{-2}$, cx2\_3:$1.65\times10^{-2}$\tabularnewline
\hline 
$Q_{3}$ & 124.01 & 62.45 & \centering{}4.95 & \centering{}$3.23\times10^{-2}$ & \centering{}$5.69\times10^{-4}$ & \centering{}cx3\_2: $1.65\times10^{-2}$, cx3\_4:$1.06\times10^{-2}$\tabularnewline
\hline 
\end{tabular}
\par\end{centering}
\caption{Calibration data of ibmq\_manila on Nov 28, 2022.cxi\_j represents
CNOT gate with control qubit i and target qubit j.\label{tab:Calibration-data-of}}
\end{table}

\begin{figure}
\centering
\includegraphics[scale=0.7]{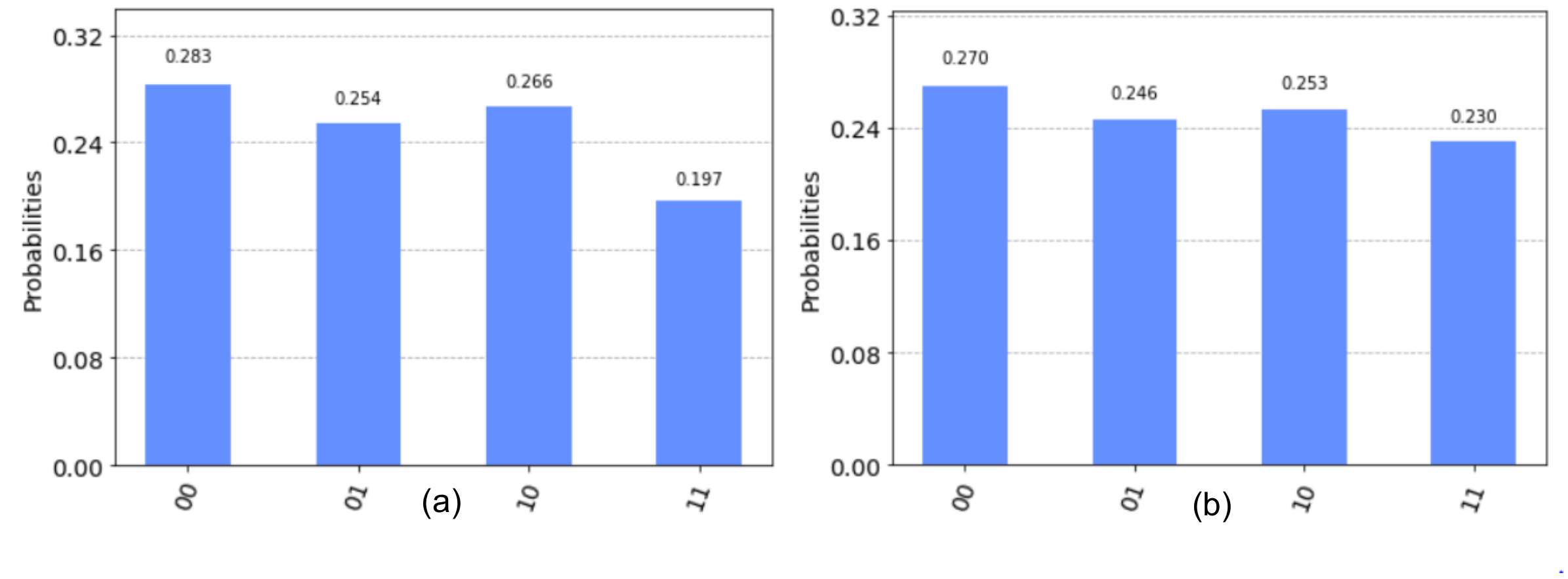}
\caption{Experimental results obtained after executing the circuit shown in (a) Fig. \ref{Comparative qcircuit}(a) and (b) Fig. \ref{Comparative qcircuit}(b) in ibmq\_manila.}
\label{Plot}
\end{figure}

The result obtained after executing the circuit is shown in Fig. \ref{Plot}. In ideal scenario, the probability of obtaining 00, 01, 10 and 11 should be equal, but here we are getting slightly different values due to error while executing the circuit. The circuit is executed for maximum number of shots (8192 shots) available on IBM Quantum experience. The number of shots represents the number of times the circuit is executed to get the final result. The fidelity between the result obtained from executing the circuit in simulator and quantum computer is calculated using $F(\sigma,\rho)=Tr\left[\sqrt{\sqrt{\sigma}\rho \sqrt{\sigma}}\right]^{2}$, where $\sigma$ representing the density matrix derived from the simulator's result and $\rho$ representing the density matrix derived from real hardware after state tomography. The average fidelity obtained for circuit shown in Fig. \ref{Comparative qcircuit}(a) and Fig. \ref{Comparative qcircuit}(b) are 86.47\% and 58.27\% respectively. The average value of fidelity is obtained by calculating it 10 times and then taking its mean. The obtained value of fidelity shows that proposed optimal scheme for broadcasting of known quantum information with two Bell state performs better than broadcasting with the cluster state.    

\section{Conclusion\label{conclusion}}
We have systematically discussed possible generalization of a set of protocols which were earlier misleadingly referred to as protocols for quantum broadcasting, but which should have been seen as broadcasting of a known quantum state to multiple receivers or just as remote state preparation at multiple ports. We have discussed effect of noise on the quantum channel used for broadcasting and found that two Bell states are more robust against noise than the cluster state. Specifically, we have shown that earlier designed schemes using 4 qubit cluster states can be realized using Bell states only. Finally, a proof of principle implementation of the proposed scheme is also performed using IBM quantum computer.

This study on broadcasting is primarily restricted to the broadcasting of the known quantum states or orthogonal quantum states which can be measured using the orthogonal basis set and considered to be known. This is so because unknown quantum states or nonorthognal quantum states can neither be cloned nor be broadcast. In this context we have already mentioned that quantum no-broadcasting theorem is a weaker version of nocloning theorem. However, it does not imply that any quantity which is nonclonable cannot be broadcast. A beautiful example is quantum Fisher information which cannot be cloned but can be broadcast even when the input states are noncommuting \cite{LSW+13}. A weaker version of broadcasting can be viewed as a situation where the output systems are not completely independent, rather they have partial overlap. If the overlap increase above a level then we can achieve a weaker version of quantum broadcasting known as quantum shared broadcasting \cite{GH07QsharedB}. In short, there are scenarios where a weaker notion of quantum broadcasting can be realized. Here we have not explored such scenarios, but we aim to extend present study to such scenarios in a future work and thus deeply explore the foundational issues related to various facets of quantum broadcasting. We conclude here with a hope that present work will be of use in future investigations on various facets of strong and weaker versions quantum broadcasting.

\section*{Acknowledgment}
Authors acknowledge the support from the QUEST scheme of Interdisciplinary
Cyber Physical Systems (ICPS) program of the Department of Science
and Technology (DST), India (Grant No.: DST/ICPS/QuST/Theme-1/2019/14
(Q80)).

\section*{Availability of data and materials}
No additional data is needed for this work.

\section*{Competing interests}
The authors declare that they have no competing interests.
\bibliographystyle{unsrt}
\bibliography{quantu-boradcast-rsp}

\end{document}